\documentclass{aa}  

\usepackage{graphicx}
\usepackage{txfonts}

\begin{document}

   \title{The effect of intermediate mass close binaries on the chemical evolution of Globular Clusters}

   \author{D. Vanbeveren\inst{1,2} \and N. Mennekens\inst{1} \and J.P. De Greve\inst{1}}
 
   \institute{Astrophysical Institute, Vrije Universiteit Brussel, Pleinlaan 2, 1050 Brussels, Belgium\\
    \email {dvbevere@vub.ac.be}
    \and Groep T - Leuven Engineering College, K.U.Leuven Association, Andreas Vesaliusstraat 13, 3000 Leuven, Belgium
    }

   \date{Received }

  \abstract
   {The chemical processes during the Asymptotic Giant Branch (AGB) evolution of intermediate mass single stars predict most of the observations of the different populations in globular clusters although some important issues still need to be further clarified. In particular, to reproduce the observed anticorrelations of Na-O and Al-Mg, chemically enriched gas lost during the AGB phase of intermediate mass single stars must be mixed with matter with a pristine chemical composition. The source of this matter is still a matter of debate. Furthermore, observations reveal that a significant fraction of the intermediate mass and massive stars are born as components of close binaries.}
   {We will investigate the effects of binaries on the chemical evolution of Globular Clusters and on the origin of matter with a pristine chemical composition that is needed for the single star AGB scenario to work.}
   {We use a population synthesis code that accounts for binary physics in order to estimate the amount and the composition of the matter returned to the interstellar medium of a population of binaries.}
   {We demonstrate in the present paper that the mass lost by a significant population of intermediate mass close binaries in combination with the single star AGB pollution scenario may help to explain the chemical properties of the different populations of stars in Globular Clusters.}
   {}

   
    \authorrunning{D. Vanbeveren et al.}  
    
    \titlerunning{The effect of binaries on the chemical evolution of Globular Clusters}
    
   \maketitle

\section{Introduction}

Globular Clusters have long been considered as examples of aggregates of stars with the same chemical composition that are born all at the same time. However, the last decade overwhelming observational evidence, spectroscopic and photometric, has been published that shows that the foregoing picture is not correct (e.g., Gratton et al. 2004; Charbonnel 2005; Carretta et al. 2009; Piotto et al. 2005, 2007; Cohen and Mel\'endez 2005; Cohen et al. 2005). 

Summarizing, the abundances of the Fe-group elements, the $\alpha$-elements and the s- and r-process elements are fairly constant from star to star but abundance variations of the light elements (C, N, O, Na, Mg and Al) have been observed in some stars of all Globular Clusters with a common pattern: C-N, O-Na and Mg-Al are anticorrelated.

In at least two Globular Clusters ($\omega$ Cen and NGC 2808) one clearly distinguishes more than one sequence among the population of hydrogen burning stars.

In Globular Clusters with more than one main sequence, the bluest ones are helium  enriched; in particular  observations of  NGC 2808 indicate that the population of hydrogen burning stars is made up of three different sequences: a normal one with Big Bang helium mass fraction (Y$\sim$0.25), an intermediate sequence with Y$\sim$0.3 and the bluest one with Y$\sim$0.35-0.4.

These observations support the self-pollution scenario where a younger generation of stars  was formed out of gas that contains the matter lost by one or more older generations, and where part of it was nuclearly processed through the CNO, NeNa and the MgAl cycles. \\

Two self-pollution scenarios have been worked out in detail. 
\begin{itemize}
\item Winds of fast rotating massive stars (WFRMS) scenario (Decressin et al. 2007): in this scenario, the younger generation is formed out of gas that was chemically enriched in the hydrogen burning zones of massive stars. CNO-processed matter is transported to the stellar surface by rotational mixing and when the massive star rotates at the critical limit these surface layers are ejected in a slow wind. However, recent studies reveal that the effect of rotation on stellar mass loss (either through a wind or trough a decretion disk) may be significantly less than the spherical, wind-like mass loss commonly assumed in evolutionary calculations  (Lovekin 2011, Krti{\v c}ka et al., 2011a, 2011b). Moreover the WFRMS scenario does not reproduce the Mg-Al anticorrelation observed in stars of Globular Clusters.\\ 
\item The Asymptotic Giant Branch (AGB) scenario (e.g., Ventura et al. 2001; Ventura \& D'Antona 2010): during the AGB phase, stars with an initial mass larger than $\sim$4 $M_{\odot}$ experience the so called `Hot Bottom Burning' = HBB (Bl\"ocker \& Sch\"onberner 1991) where the hot CNO cycle including the Ne-Na and Mg-Al chains operate at the bottom of the convective envelope. The processed matter is ejected by an AGB wind which is known to be slow enough so that this matter remains in the Globular Cluster. The AGB scenario is able to explain a large number of observations of the younger generation of stars  of Globular Clusters and it is therefore by far the better of the two, although some important issues still need to be further clarified. In particular, D'Ercole et al. (2010) implemented AGB stellar yields in a Globular Cluster chemical model and focused on the observed anticorrelations of Na-O and Al-Mg and on the helium distribution function. They concluded that correspondence with observations of the younger generation can be achieved if the nuclearly processed matter of the older generation that is returned to the interstellar medium is mixed with pristine gas that is accreted onto the Globular Cluster. The source of this pristine gas and the accretion mechanism is still a matter of debate.\\ 
\end{itemize}

Both scenarios discussed above are single star scenarios. However, there is increasing evidence that many stars are born as close binary components \footnote{close means that during the evolution at least one of both component stars will fill its Roche lobe. This will happen in most of the intermediate mass and massive binaries with orbital period up to 10 years.} and that the {\it observed} binary fraction increases with increasing primary spectral type (for reviews see e.g., Sterzik \& Durisen 2004; Kohler et al., 2006; Lada, 2006; Bouy et al., 2006; Bate, 2009). Direct observations of massive stars reveal that most of them are born as binary members (e.g., Kobulnicky \& Fryer, 2007; Sana et al. 2008; Mason et al. 2009 ). By using the adaptive optics imaging surveys of Shatsky \& Tokovinin (2002) of the association Sco OB2, Kouwenhoven et al. (2007) concluded that the primordial intermediate mass binary frequency of the aggregate was close to 100\%, 50-60\% are intermedate mass {\it close} binaries. Duquennoy \& Mayor (1991) find a Solar-type star multiplicity fraction of 0.58, the M-dwarf binary frequency = 0.42 (Fisher \& Marcy (1992) whereas probably no more than 20-25 \% of the very low mass objects (masses between 0.003 $M_{\odot}$ and 0.1 $M_{\odot}$) are binary components (for a review, see e.g., Burgasser et al., 2007). It is interesting to remark that the hydrodynamical simulations of star cluster formation of Bate et al. (2003) and Bate (2009) closely reproduce the varying observed binary frequency discussed above.  

An indirect indication that the intermediate mass close binary frequency is high comes from population synthesis  simulations. The latter reveal that in order to explain the observed supernova type Ia rate in galaxies, the intermediate mass close binary frequency must be very large, corresponding to the observations of Kouwenhoven et al. mentioned above (e.g., Han \& Podsiadlowski, 2004; Yungelson \& Livio, 2000; Ruiter et al., 2009; Mennekens et al. 2010). 

The question then naturally arises: was the initial binary frequency of the first generation of stars in the Globular Clusters similar as discussed above? The observations yield much smaller values: Albrow et al. (2001) conclude that the binary fraction in 47 Tuc is smaller than 13\% whereas Hubble Space Telescope observations of NGC 6397 yield a binary frequency $\le$ 5\%  (Davis et al., 2008).  A recent study  of Milone et al. (2011) confirms the low binary fraction in a large sample of Globular Clusters. However, the age of these clusters is very large (up to 10 Gyr and larger), and  the low mass binary frequency may have changed considerably compared to the initial one due to the dynamical interaction of stars and binaries in dense clusters as shown by Hurley et al. (2007), Fregeau et al. (2009), Chatterjee et al. (2010), Vesperini et al. (2010).  

In the present paper we will focus on the effects of massive and intermediate mass stars and binaries on the chemical evolution of Globular Clusters and we will assume that initially the Globular Cluster population is similar as the field population (similar binary frequencies, similar binary period and mass ratio distributions). Since massive and intermediate mass stars have an evolutionary lifetime $\le$ 300 Myr, the latter four studies illustrate that it is not very likely that this population is much affected by stellar dynamics.  

De Mink et al. (2009) propose non-conservative evolution of massive close binaries as the source of abundance anomalies in Globular Clusters. To illustrate their thesis, they calculate one massive binary evolutionary model, a 20 $M_{\odot}$ (primary) + 15 $M_{\odot}$ (secondary) system with an initial period of 12 days. During the Roche lobe overflow of the primary (the mass loser) the secondary (the mass gainer) spins up and reaches the critical rotation velocity. At that moment the authors assume that the mass that is lost by the loser leaves the binary taking the specific orbital angular momentum of the gainer with it. However, the critical remarks made above when the WFRMS scenario was discussed also apply here, i.e. rotation may not be responsible for large stellar mass loss and since one or both components of the massive binary will explode as a SN one may wonder if the lost matter will not be ejected from the Globular Cluster, swept up by the SN shell.

Here, we propose an alternative binary scenario. Most of the intermediate mass close binaries  (IMCB) do not experience a SN explosion whereas many lose a large amount of mass at small velocity due to Roche lobe overflow and/or the common envelope process. Part of this lost mass resembles pristine gas, part of it is helium enriched but not affected by the hot CNO burning processes. We will demonstrate in the present paper that the mass lost by a significant population of intermediate mass close binaries in combination with the single star AGB pollution scenario may explain most of the chemical properties of a second generation of stars in Globular Clusters.

\section{The evolution of intermediate mass close binaries}

Calculating the evolution of a population of intermediate mass close binaries relies on the evolution of these systems. The latter is mainly determined by three initial parameters: the mass of the primary, the mass ratio and the binary period. To explain some specific binaries, also the initial eccentricity may be important but for overall population synthesis related to the topic of the present paper (section 3) it is a second order parameter (see also Hurley et al., 2002). The evolution of IMCBs has been extensively studied since the early 60s up to the eighties by many authors (for reviews, see Paczy\'nski 1971; Plavec 1973; Van den Heuvel 1976; Thomas 1977) mainly to understand Algols, cataclysmic variables, double WD binaries and the link with Type Ia supernova, symbiotic stars, low mass X-ray binaries. De Greve and Vanbeveren (1980) compared the properties of a set of 151 theoretical binary mass exchange computations with the observed properties of Algol binaries in order to restrict the parameter space of the physics of the Roche lobe overflow. As computer power increased, much larger sets of evolutionary computations were published, e.g. Iben and Tutukov 1985, 1987; De Loore and Vanbeveren 1995; Vanbeveren et al. 1998c; Nelson and Eggleton 2001). These sets resulted in a general evolutionary scenario (see for example Hurley et al., 2002 for a general overview), a scenario that allows us to study population synthesis of intermediate mass close binaries. We discuss briefly this scenario.

The evolution of a star in a binary differs from the evolution of that star when it is single mainly due to the Roche lobe overflow (RLOF) process. Following the classification of Kippenhahn and Weigert (1967) and Lauterborn (1970) we distinguish three main phases of RLOF, which correspond to the three major expansion phases during stellar evolution: case A where the RLOF  takes place during the core hydrogen burning of the mass loser, case B where the RLOF occurs during the hydrogen shell burning phase prior to central helium burning, and case C where the RLOF begins after helium has been depleted in the core. Case B RLOF is further divided into early case B or case Br where at the onset of RLOF the envelope of the mass loser is mostly radiative and late case B or case Bc where the primary has a deep convective envelope at the beginning of the RLOF  phase. A star that already went through a first phase of RLOF during hydrogen shell burning may fill its Roche lobe for a second time during helium shell burning and perform case BB RLOF (Delgado and Thomas, 1981). 

\begin{figure*}
\centering
   \includegraphics[width=12cm]{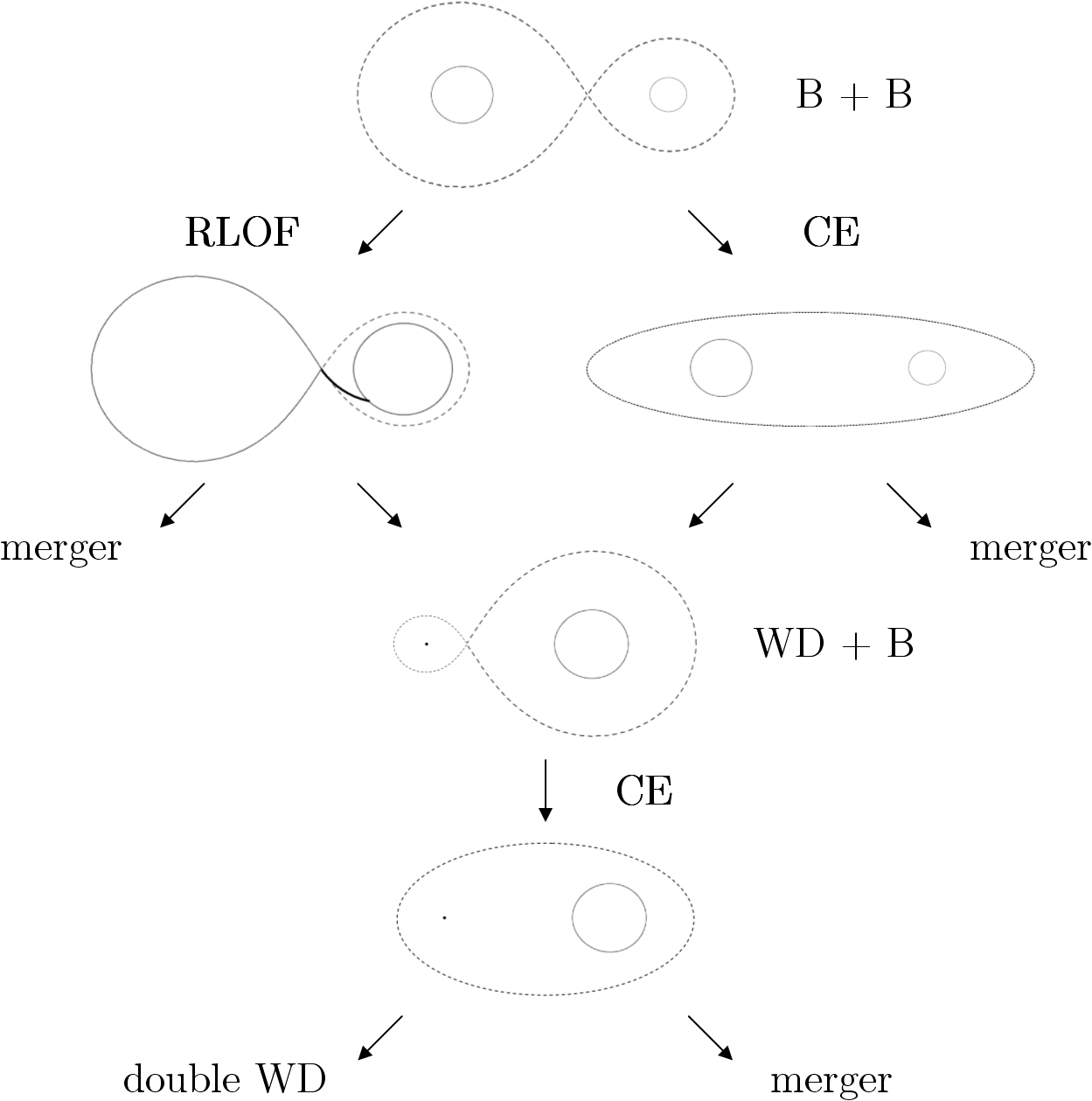}
     \caption{The overall evolution of intermediate mass close binaries with two B-type components. RLOF stand for Roche lobe overflow, CE for common envelope phase }
\end{figure*}

Figure 1 summarizes the general evolutionary scenario of intermediate mass close binaries (consisting of two B-type stars). The original most massive star (the primary) reaches its critical Roche lobe. Depending on the period of the binary the RLOF is accompanied by mass transfer or the RLOF results in a common envelope phase. In most of the binaries this phase implies the loss of hydrogen rich mass that is not affected by hot bottom burning and therefore it can serve to dilute mass that was lost by single stars during the AGB phase. When the primary has become a WD, the further evolution of the binary is governed by the secondary star. When this secondary star reaches its critical Roche lobe, a common envelope phase sets in where most of the hydrogen rich mass of the secondary is lost. Again in most of the binaries, this lost mass has not been affected by hot bottom burning and may also serve to dilute single star mass loss. Below we discuss this evolution in more detail.

\subsection{Case A and case Br}

The Brussels binary {\it evolutionary} code originates from the Paczy\'nski (1965) code. In the latter only the evolution of the donor (mass loser) was followed but it contained a detailed calculation of the mass transfer rate (imposing the condition that once the radius of the donor becomes larger than its Roche lobe, the mass loss rate is calculated so that the radius of the star equals the Roche radius). Of primary importance is the fact that this code modeled the gravitational energy loss when mass leaves the star through the first Lagrangian point, energy loss that is responsible for the luminosity drop which is typical for the evolution of the donor during its RLOF phase.  At present our code is a twin code that follows the evolution of both components simultaneously (the code has been described in detail in Vanbeveren et al., 1998a, b). The opacities are taken from Iglesias et al., (1992), the nuclear reaction rates from Fowler et al. (1975). Semi-convection is treated according to the criterion of  Schwarzschild and Harm (1958) and  convective core overshooting is included as described by Schaller et al. (1992). 

The twin code follows the evolution of the mass gainer and therefore an accretion model is essential. When the period of the binary is small enough so that the gas stream leaving the first Lagrangian point  hits the mass gainer directly, the mass gain process is treated using the formalism of Neo et al. (1977) (we call this the standard accretion model). During its RLOF, the mass loser may lose mass which has been nuclearly processed and which has a molecular weight that is larger than the molecular weight of the outer layers of the gainer. The accretion of this mass initiates mixing, a process commonly known as thermohaline convection (Kippenhahn et al., 1980).  In our code we treat this process as an instantaneous one. Note that the periods of most of the Algol binaries indicate that the mass transfer process happened and happens by direct hit and therefore it is conceivable that the standard mass gain process applies in most of these binaries. 

Mass transfer implies angular momentum transfer and the mass gainer spins up. When mass transfer proceeds via a Keplerian disk \footnote{To distinguish between direct hit or Keplerian disk we use the model developed by Lubow and Shu, 1975)}  it was shown by Packet (1981) that very soon after the onset of mass transfer, the mass gainer acquires the critical rotation velocity and therefore its evolution may be affected by rotational mixing. To simulate this mixing process, Vanbeveren et al. (1994) and Vanbeveren and De Loore (1994) introduced the accretion induced full mixing model.  In this way they were able to explain the helium and mass discrepancy in Vela X-1. The more sophisticated mass gainer models of Cantiello et al. (2007) demonstrate that the simplified models are not too bad. Remark however that rotational mixing is still heavily debated and current evolutionary models of single stars including rotational mixing do not seem to explain the observations (Hunter et al.,  2008). In our code we can switch off this process (e.g., mass accretion is treated in the standard way in all the case A/Br binaries). 

An interesting conclusion resulting from all caseA/Br intermediate mass close binary evolutionary computations done in the past is that the overall evolution of the mass loser is largely independent from the details of the RLOF process, from the initial period of the binary, from the mass of its companion and from the initial metallicity. The RLOF stops when both stars merge (see section 2.6) or in case A binaries when the mass loser reaches the main sequence hook (end of core hydrogen burning) or in case Br binaries when the loser has lost most of its hydrogen rich layers and helium starts burning in the core. The evolution of the mass gainer however, depends on the details of the RLOF, on the adopted accretion model and on the amount of accreted mass. Moreover, the simultaneous evolutionary computations of caseA/Br binaries reveal that in most of them both stars come into contact during RLOF. It is not unreasonable to assume that mass will leave the binary when a contact system is formed but since the physics of contact is poorly constrained it is uncertain how much mass will leave the binary. This uncertainty may imply an important uncertainty for all applications of intermediate mass close binary evolution. For this reason De Loore and Vanbeveren (1995) and Vanbeveren et al. (1998) calculated several thousands of evolutionary tracks of mass gainers for different initial chemical compositions, for different accretion efficiency characterized by $\beta = \left|\dot{M_{2}} / \dot{M_{1}}\right|$, which is the fraction of matter lost by the donor (subscript 1) that is accreted by the gainer (subscript 2). This makes it possible to study the effect of the uncertainty of the accretion efficiency on population studies, like the one of the present paper.

If $\beta < 1$, and thus mass is lost from the system, it is necessary to make an assumption about how much angular momentum this lost mass takes with it. This quantity is obviously dependent on the physical model of how this matter escapes from the system, and is critical for the orbital period evolution during the mass transfer phase. Our standard model is that matter escaping will do so via the second Lagrangian point $L_2$, from where it forms a circumbinary disk (van den Heuvel, 1993). A 'bare-minimum' for the radius of this disk is obviously equal to the distance from $L_2$ to the center of mass. It has been shown by De Donder \& Vanbeveren (2004) that this value varies only very slightly during the mass transfer phase, resulting in an average $\eta$ (= diameter of the disk/orbital separation) of 1.3. However, Soberman et al. (1997) concluded that circumbinary disks are stable (e.g., the matter in the disk will not have the tendency to fall back towards the binary) only when their radii are at least a few times the binary separation and they propose $\eta$ = 2.3 (this value of $\eta$ is the standard value used in the Brussels {\it 'evolutionary'} and {\it 'population synthesis'} code). Under the assumption that  $\eta$ is constant the period change is given by (see De Donder \& Vanbeveren 2004)\\

for $0 < \beta < 1$:

\begin{equation}
\frac{P_\mathrm{f}}{P_\mathrm{i}} = \left(\frac{M_{1\mathrm{f}}+M_{2\mathrm{f}}}{M_{1\mathrm{i}}+M_{2\mathrm{i}}}\right)\left(\frac{M_{1\mathrm{f}}}{M_{1\mathrm{i}}}\right)^{3\left[\sqrt{\eta}\left(1-\beta\right)-1\right]}\left(\frac{M_{2\mathrm{f}}}{M_{2\mathrm{i}}}\right)^{-3\left[\sqrt{\eta}\frac{1-\beta}{\beta}+1\right]};
\end{equation}

for $\beta = 0$:

\begin{equation}
\frac{P_\mathrm{f}}{P_\mathrm{i}} = \left(\frac{M_{1\mathrm{f}}+M_{2\mathrm{f}}}{M_{1\mathrm{i}}+M_{2\mathrm{i}}}\right)\left(\frac{M_{1\mathrm{f}}}{M_{1\mathrm{i}}}\right)^{3\left(\sqrt{\eta}-1\right)}\mathrm{e}^{3\sqrt{\eta}\left(\frac{M_{1\mathrm{f}}-M_{1\mathrm{i}}}{M_{2\mathrm{i}}}\right)}.
\end{equation}

In these formulae, subscripts i and f indicate the situation before and after RLOF. By using the formulae given above it can readily be checked that our standard model predicts that the binary period decreases significantly during a non-conservative RLOF and this implies that the two binary components may merge. We discuss mergers in a separate subsection.

An often made assumption is that matter escapes with the specific orbital angular momentum of the gainer star. This may apply when matter escapes from the gainer as a symmetrical wind (spherical or bipolar for which some evidence is found in a few Algol systems). In this case the period change is given by\\

for $0 < \beta < 1$:

\begin{equation}
\frac{P_\mathrm{f}}{P_\mathrm{i}} = \left(\frac{M_{1\mathrm{f}}+M_{2\mathrm{f}}}{M_{1\mathrm{i}}+M_{2\mathrm{i}}}\right)^{-2}\left(\frac{M_{1\mathrm{f}}}{M_{1\mathrm{i}}}\right)^{-3}\left(\frac{M_{2\mathrm{f}}}{M_{2\mathrm{i}}}\right)^{-3\left(\frac{1-\beta}{\beta}+1\right)}.
\end{equation}

for $\beta = 0$:

\begin{equation}
\frac{P_\mathrm{f}}{P_\mathrm{i}} = \left(\frac{M_{1\mathrm{f}}+M_{2\mathrm{f}}}{M_{1\mathrm{i}}+M_{2\mathrm{i}}}\right)^{-2}\left(\frac{M_{1\mathrm{f}}}{M_{1\mathrm{i}}}\right)^{-3}\mathrm{e}^{3\frac{M_{1\mathrm{f}}-M_{1\mathrm{i}}}{M_{2\mathrm{i}}}}.
\end{equation}

To illustrate the effect of the adopted mass loss model during case A/Br RLOF on the results of the present paper we will present computations when the latter assumption applies and when the standard model is used. 

We end this subsection with two remarks. 

(1) When the mass transfer in a case A/Br intermediate mass close binary is quasi-conservative, the mass of the gainer may become larger than the minimum mass of a supernova type II event to occur.  This SN II is preceded by the common envelope phase of the WD+B type binary (Figure 1 and subsection 2.5). Since the mass lost from the binary during this common envelope phase may be ejected from the Globular Cluster by this supernova, our results discussed in section 4 do not account for this lost mass. 

(2) Evolutionary computations of binaries  reveal that although an intermediate mass star shrinks in response to the loss of matter, it may not shrink fast enough to compete with the rapid shrinkage of the Roche radius when initially the binary has a mass ratio q $<$ 0.2. This is a forciori true when the star has passed the point on the main sequence when the central hydrogen content Xc $\sim$ 0.1. It is therefore conceivable that in most of these binaries the secondary is engulfed by the primary during the RLOF and both components merge. We evolve such systems by means of the common envelope formalism (see subsection 2.3).

\subsection{Case BB}

After case A/Br mass transfer, the remnant central helium burning star may grow to giant dimensions during He shell burning and fill its Roche lobe for a second time, initiating a phase of case BB mass transfer (Delgado and Thomas, 1981; Habets, 1986a, b; Avila Reese, 1993). The star loses its remaining hydrogen layers and most of its helium layers on top of the helium burning shell. The mass loss rates of the Roche lobe filling component during case BB RLOF are much smaller than during case A/Br RLOF (Dewi et al., 2002). This implies that if during case BB RLOF the mass gainer is a normal star, mass transfer will proceed conservative. The conservative model is also the model that is used in our evolutionary calculations and in our population synthesis simulations discussed in section 4.

\subsection{Case Bc}

Case Bc  RLOF is unstable which means that the expansion of the donor will not be abated by mass loss, and its outer layers will eventually engulf the other star, resulting in a common envelope phase. The viscous friction of both stellar cores rotating within the shared atmosphere will result in a decrease in orbital period. The orbital energy thus lost will be partially converted into kinetic energy used to expell the common envelope. If this conversion is sufficiently efficient, the entire envelope may be ejected and a binary emerges with smaller orbital period. Otherwise, the two cores will merge before this can happen. Because of the very short timescale of a common envelope phase, it is assumed that the gainer will not accrete an appreciable amount of mass, hence $\beta = 0$. The orbital separation evolution is dictated by the $\alpha$-formalism by Webbink (1984):

\begin{equation}
\frac{M_{1\mathrm{i}}\left(M_{1\mathrm{i}}-M_{1\mathrm{f}}\right)}{\lambda R_{\mathrm{Roche}}} = \alpha \left(\frac{M_{1\mathrm{f}}M_{2\mathrm{i}}}{2A_\mathrm{f}}-\frac{M_{1\mathrm{i}}M_{2\mathrm{i}}}{2A_\mathrm{i}}\right),
\end{equation}

\noindent where $R_{\mathrm{Roche}}$ is the Roche radius of the donor, $\lambda$ is determined by the density structure of its outer atmosphere and on its internal energy that can help to expel the common envelope (see Dewi and Tauris, 2000 for a detailed description and computation), and $\alpha$ is describing the efficiency of the energy conversion. The observations of post common envelope binaries of Zorotovic et al. (2010) seem to indicate that on average these systems can be explained by assuming an average $\alpha\lambda$ = 0.1-0.2 (see also Davis et al., 2010). The stellar structure computations of Dewi and Tauris (2000) also reveal such small values for many red giants although values around 1 can not be excluded for AGB stars with loosely bound envelopes. We will present population synthesis computations for $\alpha\lambda$ = 0.1, 0.5 and 1. In order to illustrate the uncertainty caused by the assumption $\alpha\lambda$ =  constant during the common envelope phase of all binaries, we also calculated a model where $\alpha$ = a constant but $\lambda$ is computed using the tabulated results of Dewi and Tauris. 

Similarly to the RLOF in case Br binaries, the common envelope phase in case Bc binaries stops when the donor has lost most of its hydrogen rich layers and helium starts burning in its core or when the two components merge (see subsection 2.6). Case Bc donors are hydrogen shell burning stars and they have a very similar structure as case Br donors. When the binary survives the common envelope phase it can therefore be expected that the donor remnant has a structure that is very similar to the structure of a case Br donor remnant.  In our population synthesis simulations (section 4) we assume that the case Bc donor remnants are the same as the case Br donor remnants. After the common envelope phase the case Bc remnants evolve in a similar way as case Br remnants so that they may perform a case BB RLOF during their He shell burning phase as well. 

\subsection{Case C}

Similarly as in case Bc binaries, case C RLOF is unstable and therefore the common envelope scenario is used to compute the binary period evolution. Also here the RLOF/common envelope will stop when all hydrogen rich layers are removed. To calculate how much mass is left (as well as the chemistry of these layers) we use the single star tracks of Schaller et al. (1992) and the AGB evolutionary calculations of van den Hoek and Groenewegen (1997). The evolution of AGB binary components in the pre-common envelope phase of case C systems is affected by stellar wind mass loss for which van den Hoek and Groenewegen use the Reimers (1975) formalism. Stellar wind mass loss decreases the total mass lost during the common envelope phase and it affects the overall orbital period evolution of the binary, e.g. the period variation due to stellar wind is given by

\begin{equation}
\frac{P_\mathrm{f}}{P_\mathrm{i}} = \left(\frac{M_{1\mathrm{f}}+M_{2\mathrm{f}}}{M_{1\mathrm{i}}+M_{2\mathrm{i}}}\right)^{-2}.
\end{equation}

In this formula, subscripts i and f indicate the situation at the beginning of the stellar wind phase and at the end just before the onset of the common envelope phase.

\subsection{The evolution of a binary with a White Dwarf component and a normal star }

During the second mass transfer phase (when the originally most massive star has become a WD and the other is now filling its Roche lobe), mass transfer occurs towards the small surface of a WD. Furthermore, most of these binaries have an extreme mass ratio and therefore, it can be expected that such mass transfer will always be dynamically unstable, resulting in a common envelope-induced spiral-in of the WD into the non-degenerate star's outer layers. One exception to this is when the combination of the mass of the non-degenerate star and the orbital period at the onset of mass transfer falls within the zones identified by Hachisu et al. (2008) as progenitors of single degenerate type Ia supernovae. In that case, it is assumed that accretion towards the WD will occur in a stable way (due to a stabilizing wind from the WD) until it reaches the Chandrasekhar limit and explodes in such an event. Type Ia supernova yields are obviously not included in this study, as they are not ejected through slow winds. Similarly as in the caseA/Br/Bc/C binaries discussed above, the common envelope phase here will stop when the normal star has lost all the hydrogen rich layers leading to the formation of a double WD binary. To calculate the mass lost by the normal star and the chemistry of this lost mass we use the same procedure as outlined in the previous subsections. Note that since the common envelope phase results in a very significant shrinking of the orbital period, a WD + normal star binary can merge. 

As mentioned already in subsection 2.1, when the WD + normal star binary originated from a quasi-conservative case A/Br progenitor, the mass of the normal star may be larger than the minimum mass for a supernova type II. In our population simulations, we assume that in this case the mass lost during the common envelope of the WR + normal star binary will be ejected from the Globular Cluster and therefore this lost mass is not included in the simulations discussed in section 4.

\subsection{Mergers }

In a significant fraction of intermediate mass close binaries, the two components will merge at some point during their evolution, before a double WD binary is obtained. This merger can occur either during an episode of stable Roche lobe overflow (with the evolution of the orbit described by Eq. (1), (2), (3) or (4)), or during a common envelope phase (Eq. (5)). The criterion that we use to determine whether a given system will survive a mass transfer episode is to compare the theoretical stellar equilibrium radii of both stars after mass transfer (determined from their masses at that time) with the corresponding Roche radii. \footnote{During the Roche lobe overflow or common envelope phase the mass loser is not in thermal equilibrium, however,  when the mass loss phase stops, e.g. when the star has lost most of its hydrogen rich layers, it regains its equilibrium very rapidly. The same applies for the mass gainer in a case Br binary. During mass accretion the star is out of equilibrium but when the accretion stops it quickly restores its equilibrium.} When at least one of the equilibrium radii is larger than the corresponding Roche radius the system merged. The further evolution of a merger will be that of a (possibly exotic) single star, and its ejecta are no longer included in this study. The code however does include, in the case of systems that merge during a mass transfer phase, the matter lost through nonconservative Roche lobe overflow or common envelope evolution up to that moment. This is achieved by determining how much mass can be lost by the donor star before the above criterion for survival no longer holds.


\section{The Brussels population synthesis  code}

To compute the total mass lost by a population of intermediate mass close binaries, the overall evolutionary scenario discussed in the previous section has to be combined with an initial mass function of binary primaries, a binary mass ratio and orbital period distribution.  In our code the primary masses follow a Kroupa et al. (1993) type initial mass function, normalized between 0.1 and 120 $M_{\odot}$. We adopt a standard binary period distribution (see e.g. Abt 1983) which is flat in log period, with initial orbital periods between one day and ten years. We consider three different mass ratio ($q$) distributions (with $0 < q < 1$): a flat one, a distribution favoring binaries with small mass ratio (Hogeveen 1992), and a distribution favoring binaries with mass ratio close to one (Garmany et al. 1980). The computations are performed for $\beta$ between 0 (all mass lost by the primary during a dynamically stable RLOF leaves the binary as described in Sect. 2) and 1 (conservative mass transfer during a dynamically stable RLOF). Two angular momentum (AM) loss mechanisms are considered (see Sect. 2): AM of the second Lagrangian point ("$L_2$") with $\eta$ = 2.3 (our standard model)  and specific orbital AM of the gainer ("O").  To illustrate the effect of the adopted value of $\eta$ in the standard model we also made population synthesis simulations with $\eta$ = 1.

The common envelope phases are treated by using the Webbink (1984) formalism with different values of the parameter $\alpha\lambda$, as explained in Sect. 2.  We will present population synthesis computations for $\alpha\lambda$ = 0.1, 0.5 and 1. In order to illustrate the uncertainty caused by the assumption $\alpha\lambda$ = a constant during the common envelope phase of all binaries, we also calculated a model where $\alpha$ = a constant but $\lambda$ is computed using the tabulated results of Dewi and Tauris.

From the evolutionary scenario described in the previous section we know the chemistry of the layers lost by intermediate mass close binaries during their RLOF . Our population synthesis code calculates the total mass lost by a population of intermediate mass close binaries (called $\Delta M$, and expressed as a fraction of the mass contained in these systems) and the chemistry of this lost mass. Part of this lost mass is not affected by CN, CNO and/or Ne-Na and Mg-Al reactions. As far as C, N, O, Ne, Na, Mg, Al is concerned this is gas with a pristine chemical composition and we therefore call this mass $\Delta M_{pris}$. Part of it is affected by the CN or CNO cycles, e.g, this gas has CN or CNO equilibrium abundances, it is He enriched but it is not affected by the Ne-Na or by the Mg-Al reactions. Therefore this gas is helium enriched gas with CN or CNO equilibrium abundances but with a pristine chemical composition as far as Ne, Na, Mg and Al is concerned (we denote this as $\Delta M_{pris+He}$). We have decided to consider separately the mass lost in a binary where at least one of the components was an AGB star at the onset of Roche lobe overflow (a case C binary). We make a distinction between case C binaries where one of the components fills its Roche lobe during the early AGB (EAGB) phase and those where the Roche radius is reached during the TPAGB phase. The convective envelope of a star lost during the EAGB phase has not yet been affected by HBB and the chemical composition is therefore pristine as far as Ne, Na, Mg and Al is concerned, the CNO composition is affected by CN and CNO reaction rates and it is enriched with He (we note this as $\Delta M_{EAGB}$. The mass lost during the TPAGB phase however, may have been affected by hot bottom burning, thus by the Ne-Na and Mg-Al chains. We are aware that this hot bottom burning is a function of mass (e.g., not all intermediate mass AGB stars undergo hot bottom burning) and the physics of this process in not yet very well constrained (e.g., Karakas and Lattanzio, 2007; Gil-Pons et al., 2007; Izzard et al., 2007; Ventura and D'Antona, 2010). We therefore simply compute the mass lost during the TPAGB phase without further details (we use $\Delta M_{TPAGB}$). We like to remark here that the TPAGB chemical yields of case C binary components may be significantly different from TPAGB single star yields. Single stars go through the full TPAGB phase whereas depending on the binary period the RLOF will stop the TPAGB phase of the binary component before this phase is finished, e.g. interaction with a binary companion will affect most of the TPAGB chemistry. In the present paper we did not account for this effect since we focus on the amount of mass lost by a binary population that is NOT affected by hot bottom burning, matter that may help to explain the observed Na-O and Al-Mg anti-correlations.

\section{Results and discussion}

To illustrate how intermediate mass close binaries eject mass in the interstellar medium, Table 1 shows a number of numerical examples concerning  mass loss from binaries due to the Roche lobe overflow and common envelope processes. The first part of the table deals with a system with initial ZAMS masses of 6.0 and 5.4 M$_{\odot}$. It shows, depending on the range of initial orbital periods, the different evolution regimes possible for such systems (starting with the nature of the mass transfer episode caused when the primary fills its Roche lobe, indicated by MT1 in the table). 

\begin{table*}
\caption{Examples of  ejecta for various initial period ranges. First section is for a 6.0+5.4 M$_{\odot}$ system and with $\beta=1$ (second section is with $\beta=0.5$). Third section is for a 6.0+3.6 M$_{\odot}$ system with $\beta=1$ (fourth section is with $\beta=0.5$).}
\label{table3}
\centering
\begin{tabular}{c c c c c c c c c c c }
\hline
$P_{min}$ & $P_{max}$ & MT1 & time & total & He & TPAGB & MT2 & time & total & He  \\
(d) & (d) & type & (Myr) & ej. (M$_{\odot}$) & ej. (M$_{\odot}$) & ej. (M$_{\odot}$) & type & (Myr) & ej. (M$_{\odot}$) & ej. (M$_{\odot}$)  \\
\hline
1.0 & 75 & A/Br & 72 &  &  &  & SNII & & & \\
75 & 150 & Bc M & 72 & 4.0-4.6 & 1.0-1.3 &  & none & & & \\
150 & 240 & Bc & 72 & 4.6 & 1.3 &  & B M & 79 & 0.9-1.1 & 0.2-0.3  \\
240 & 310 & C M & 78 & 4.9-5.0 & 1.4-1.5 &  & none & & &  \\
310 & 1300 & C & 78 & 5.0 & 1.5 &  & C & 88 & 4.6 & 1.5  \\
1300 & 3700 & C & 78 & 5.0 & 1.5 & 5.0 & C & 88 & 4.6 & 1.5  \\
\hline
1.0 & 3.8 & A/Br M & 72 & 0.3-2.0 & 0.1-0.5 &  & none & & & \\
3.8 & 75 & Br & 72 & 2.3 & 0.7 &  & A/B M & 92-102 & 0.5-2.7 & 0.2-0.8  \\
\hline
1.0 & 11 & A/Br & 72 &  &  &  & B M & 106-107 & 2.3-6.2 & 0.7-1.7 \\
11 & 23 & Br & 72 &  &  &  & B & 107 & 7.0 & 2.4 \\
23 & 71 & Br & 72 &  &  &  & C & 151 & 6.7 & 2.4  \\
\hline
1.0 & 11 & A/Br M & 72 & 0.3-2.2 & 0.1-0.6 &  & none & & & \\
11 & 71 & Br & 72 & 2.3 & 0.7 &  & A/B M & 140 & 0.5-1.5 & 0.2-0.5 \\
\hline
\end{tabular}
\end{table*}

If the initial orbital period is below 75 days, the primary will after 72 Myr initiate a case A or radiative case Br Roche lobe overflow, resulting in stable mass transfer. In the first part of the table, it is assumed that this phase is conservative ($\beta=1$), and thus no mass is lost from the system during this episode. The secondary accretes enough matter so that it later explodes as a core-collapse SN. This means that although the secondary loses its hydrogen rich layers during its common envelope phase, it is expected that this mass will be ejected from the Globular Cluster and it is therefore not included in the table. 

With an initial orbital period of 75 to 150 days, the outer layers of the primary will have become deeply convective by the time it fills its Roche lobe. The system therefore undergoes an unstable case Bc, resulting in a (non-conservative) common envelope phase. Our simulations reveal that  the system merges (indicated in the table by the letter M). Depending on the orbital period, this will happen after the primary has ejected 4.0 to 4.6 M$_{\odot}$, of which 1.0 to 1.3 M$_{\odot}$ helium, through the common envelope. 

When the orbital period is larger than 150 days, the system will not merge during the common envelope process. It will undergo a second mass transfer episode (MT2 in the table) initiated by the secondary at 79 Myr. This will be a case B resulting in spiral-in (as always when a WD is involved) during which the system merges. Before the merger, an additional 0.9 to 1.1 M$_{\odot}$, of which 0.2 to 0.3 M$_{\odot}$ helium, is ejected. 

With an initial orbital period between 240 and 310 days, the primary will fill its Roche lobe after core helium depletion, and thus initiate (unstable) case C mass transfer at 78 Myr, which ends because the stars merge. Again depending on the exact period, 4.9 to 5.0 M$_{\odot}$ will be ejected beforehand, of which 1.4 to 1.5 M$_{\odot}$ helium. 

With an initial period above 310 days, the system survives the common envelope. When the secondary fills its Roche lobe (which is a case C at 88 Myr) another 4.6 M$_{\odot}$ of which 1.5 M$_{\odot}$ helium is ejected. The reason that a non-merging primary case C ejects more matter than a non-merging primary case Bc is that in the latter case, part of the primary mass loss occurs during the so called case BB mass transfer, which is always considered to be conservative and does thus not cause mass loss from the system. In a case C however, the primary is taken to lose its entire envelope in one episode of common envelope evolution. 

For systems with an initial period above 1300 days, the same values hold as in the previous paragraph, but since the primary then only fills its Roche lobe when it has become a TPAGB star, the ejecta during the first mass transfer phase is then considered TPAGB enriched. It should be noted that regardless of the orbital period, TPAGB enrichment during the second mass transfer episode is negligible in all systems.

The second part of Table 1 revisits the case A/Br of systems below 75 days, but now under the assumption that stable mass transfer occurs semi-conservatively ($\beta=0.5$). In that case, due to the angular momentum loss, systems below 3.8 days will merge. Obviously, the non-conservatism also means that matter is ejected beforehand, 0.3 to 2.0 M$_{\odot}$ of which 0.1 to 0.5 M$_{\odot}$ helium. Above 3.8 days, the system survives and this amount increases to 2.3 M$_{\odot}$ of which 0.7 M$_{\odot}$ helium, plus 0.5 to 2.7 M$_{\odot}$ of which 0.2 to 0.8 M$_{\odot}$ helium during the mass transfer episode of the secondary. This episode results in a merger and happens between 92 and 102 Myr, depending on whether accretion during the first mass transfer occurs through direct impact or an accretion disk (the latter causing accretion induced full mixing, rejuvinating the secondary and thus delaying its evolution). 

The last two parts of Table 1 again focus on the period range where mass transfer is stable (with $\beta=1$ in part 3 and $\beta=0.5$ in part 4), but now for a companion with an initial mass of 3.6 M$_{\odot}$. In the conservative case, no matter is lost during the first mass transfer phase, but depending on the initial orbital period, the second mass transfer phase is either a case B merger, a survivable case B, or a survivable case C. Here, more matter is lost in the case B than in the case C as the core has grown larger in the latter case, and there is no possibility for case BB towards a WD to compensate for this. In the semi-conservative case, systems below 11 days will merge during the first mass transfer phase, while those above will survive and cause a second (merging) case A or B mass transfer phase.

Table 2 summarizes our population synthesis simulations for a population consisting of 100\% intermediate mass close binaries with primary mass $M_1$ between 3 $M_{\odot}$ and 10 $M_{\odot}$ and initial chemical composotion (X, Y, Z) = (0.26, 0.24, 0.0001). The evolution of every binary in the population code is followed as illustrated in Table 1.

\begin{table*}
\caption{Results obtained with the PNS code for a population of 100\% IMCBs. See text for definition of symbols.}
\label{table1}
\centering
\begin{tabular}{c c c c c c c c c c}
\hline
$q$-distr. & $\beta$ & $\alpha\lambda$ & AM loss & $\Delta M$ & $\frac{\Delta M_{pris}}{\Delta M}$ & $\frac{\Delta M_{pris+He}}{\Delta M}$ & $\frac{\Delta M_{EAGB}}{\Delta M}$ & $\frac{\Delta M_{TPAGB}}{\Delta M}$ & Y \\
\hline
flat & 1.0 & 1.0 & $L_2(\eta=2.3)$ & 40\% & 23\% & 22\% & 47\% & 8\% & 0.30\\
flat & 0.5 & 1.0 & $L_2(\eta=2.3)$ & 34\% & 43\% & 7\% & 40\% & 9\% & 0.28\\
flat & 0.0 & 1.0 & $L_2(\eta=2.3)$ & 31\% & 49\% & 2\% & 39\% & 10\% & 0.27\\
flat & 1.0 & 0.5 & $L_2(\eta=2.3)$ & 33\% & 19\% & 18\% & 53\% & 9\% & 0.30\\
flat & 0.0 & 0.5 & $L_2(\eta=2.3)$ & 27\% & 46\% & 1\% & 41\% & 11\% & 0.26\\
flat & 1.0 & 0.1 & $L_2(\eta=2.3)$ & 20\% & 14\% & 11\% & 62\% & 14\% & 0.31\\
flat & 0.5 & 0.1 & $L_2(\eta=2.3)$ & 18\% & 45\% & 5\% & 36\% & 15\% & 0.26\\
flat & 1.0 & "Dewi" & $L_2(\eta=2.3)$ & 28\% & 19\% & 12\% & 58\% & 11\% & 0.30\\
flat & 1.0 & 1.0 & O & 43\% & 24\% & 23\% & 45\% & 7\% & 0.30\\
flat & 0.5 & 1.0 & O & 53\% & 35\% & 18\% & 41\% & 6\% & 0.29\\
flat & 1.0 & 1.0 & $L_2(\eta=1)$ & 41\% & 24\% & 23\% & 46\% & 8\% & 0.30\\
flat & 0.5 & 1.0 & $L_2(\eta=1)$ & 44\% & 38\% & 20\% & 34\% & 7\% & 0.28\\
Garmany & 1.0 & 1.0 & $L_2(\eta=2.3)$ & 39\% & 22\% & 30\% & 40\% & 8\% & 0.30\\
Garmany & 0.0 & 1.0 & $L_2(\eta=2.3)$ & 30\% & 50\% & 2\% & 38\% & 10\% & 0.27\\
Hogeveen & 1.0 & 1.0 & $L_2(\eta=2.3)$ & 41\% & 29\% & 19\% & 43\% & 8\% & 0.29\\
Hogeveen & 0.0 & 1.0 & $L_2(\eta=2.3)$ & 35\% & 46\% & 2\% & 42\% & 10\% & 0.27\\
\hline
\end{tabular}
\end{table*}

Our simulations lead to the following conclusions.

\begin{itemize}

\item Independent from the adopted initial mass ratio distribution, a first generation of intermediate mass close binaries returns 20-40\% of its mass. Typically 80-90\% of this mass is gas that has not been affected by the Ne-Na and Mg-Al reactions. Part of this gas is He-enriched during core hydrogen burning.  10-20\% of the returned mass was ejected by binary components that went through a TPAGB phase and therefore this mass may have an AGB type composition, but as remarked above this composition will be different from single stars.\\ 
\item We started with a population of binaries with zero age main sequence helium abundance = 0.24. The average helium abundance of the returned mass typically equals 0.26- 0.3. \\ 
\item The above fractions depend only slightly on the adopted value of $\beta$ during the Roche lobe overflow in a case A/Br binary.  \\
\item The above fractions differ by more than a factor 2 when a model with $\alpha\lambda$ = 1 is compared with a model with $\alpha\lambda$ = 0.1. \\ 
\item The results given in Table 2 are calculated adopting the accretion induced full mixing process when the stable mass transfer proceeds via a Keplerian disc (section 2). The results differ by less than 3\% when this full mixing process is switched off.\\
\item The result with a variable $\alpha\lambda$ (denoted by Dewi) is similar to the result with a constant $\alpha\lambda$ = 0.25.

\end{itemize}

Figure 2 shows the ejecta of a population of intermediate mass close binaries as a function of time since starburst. It includes the total ejected mass, the ejected mass in helium (both primordial and newly synthesized) and the ejected mass of TPAGB-enriched matter. It has been calculated using the models with a flat $q$-distribution, $\beta = 1$, $\alpha\lambda = 1$ and 0.1, $\eta = 2.3$ (meaning mass loss through $L_2$). Note that with this choice of parameters the total amount of ejected material has been maximized but the plots for any other combination of parameters included in Table 1 would look very similar in overall shape. Most of the mass is ejected between 30 and 150 Myr. The two simulations in Figure 2 show two peaks. The first peak is due to the ejecta of the mass losers in the synthesized population of intermediate mass close binaries, the second peak is due to the ejecta of mass gainers. This is further illustrated in Figure 3 where we show the temporal evolution of the ejecta of one of the models of Figure 2 but we separated the contributions of the mass losers and the mass gainers. It can thus be expected that the second peak will be smaller when the stable Roche lobe overflow in case Br binaries will be highly non-conservative. Figure 3 illustrates indeed that the peak almost disappears in a simulation where it is assumed that  $\beta = 0$.

\begin{figure*}
\centering
   \includegraphics[width=12cm]{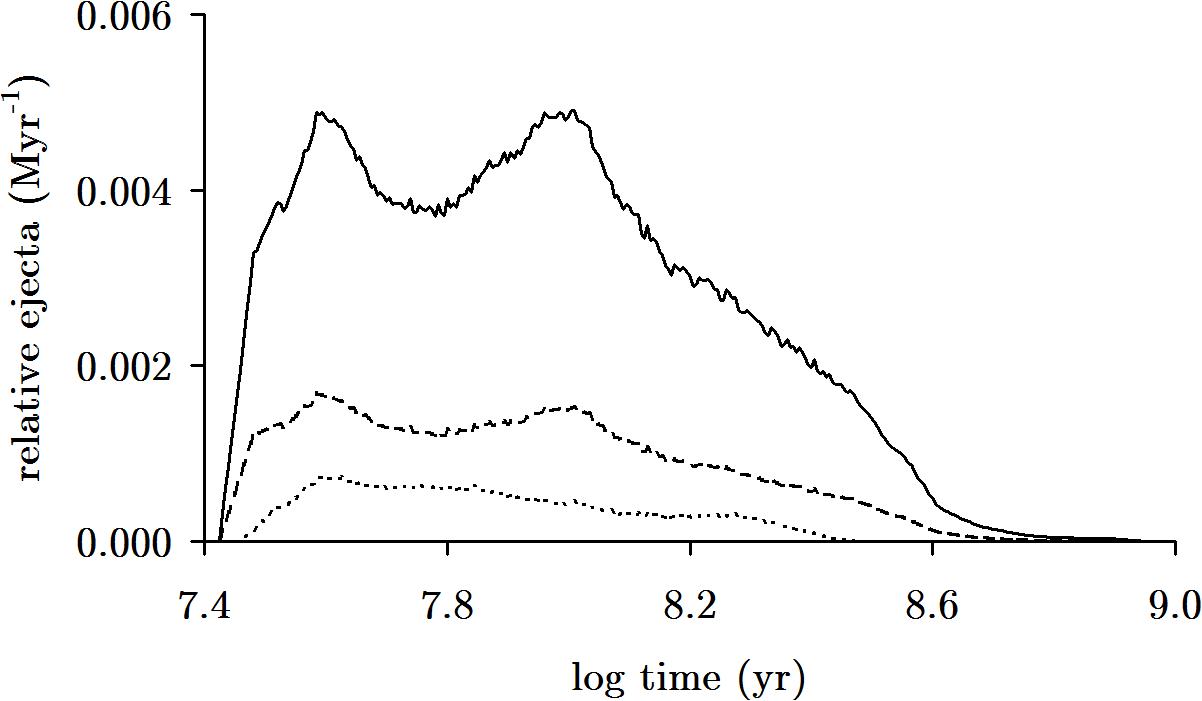}
     \includegraphics[width=12cm]{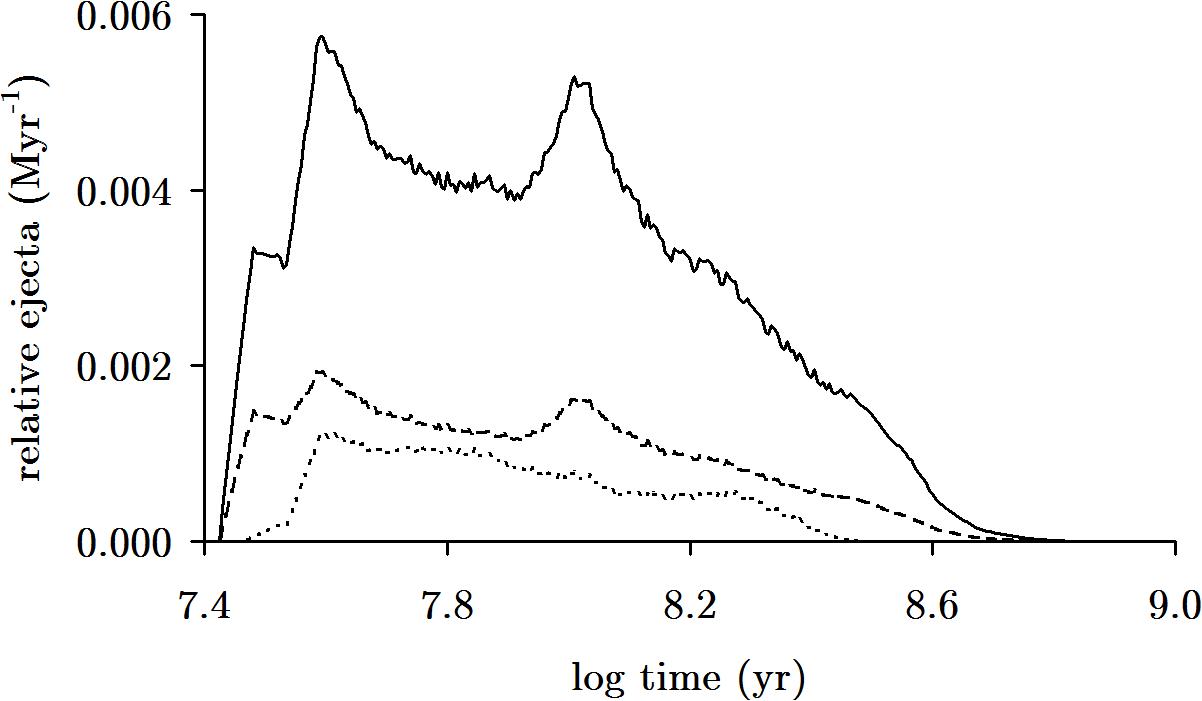}
     \caption{Ejecta per Myr of a population of 100\% intermediate mass close binaries as a function of time since starburst, normalized to the total mass ejected. Displayed are total ejecta (solid), ejected He (dashed) and ejected TPAGB-enriched material (dotted). Both figures describe a model with a flat $q$-distribution, $\beta = 1$, $\eta = 2.3$. The top (respectively bottom) figure is calculated with $\alpha\lambda = 1$ (respectively  0.1). }
\end{figure*}

\begin{figure*}
\centering
   \includegraphics[width=12cm]{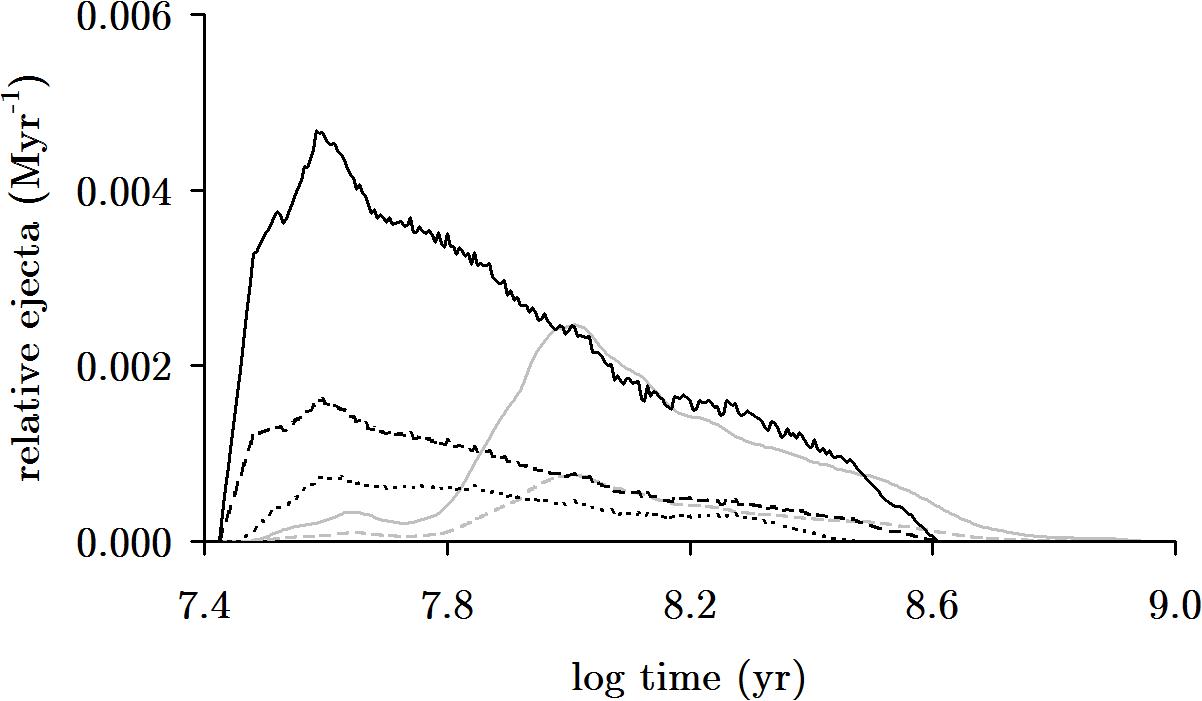}
     \caption{Similar as Figure 2a but we separately consider the contribution of the losers (black) and the gainers (gray). }
\end{figure*}

\begin{figure*}
\centering
   \includegraphics[width=12cm]{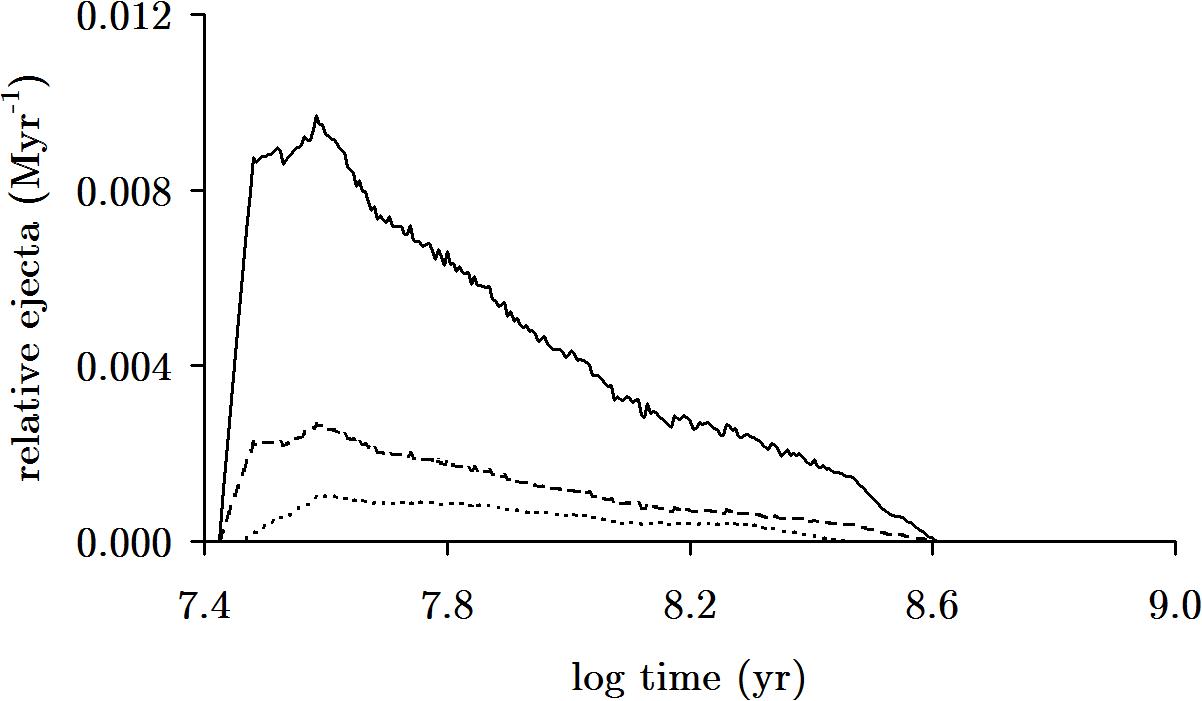}
     \caption{Similar as Figure 2a but $\beta=0$. }
\end{figure*}

As a byproduct our population synthesis simulations predict the population of mergers (note that mergers lose mass during the merger process, subsection 2.6, and this lost mass  is significant compared to the total mass lost by a population). Table 3 shows the total (pre double WD) merger rate in intermediate mass close binaries, as well as the fraction of mergers that respectively take place during stable Roche lobe overflow, common envelope phase (both during the first mass transfer phase), and during the second mass transfer phase (involving a WD spiral-in). We notice that the majority of the intermediate mass close binaries will experience a merger event. These mergers will further evolve as single stars but as single stars which are completely different from canonical single stars. A detailed study on the evolution of such mergers (especially mergers where a WD is involved) has not been done yet and therefore it is unknown how these mergers will affect the chemical history of a Globular Cluster. But since many are predicted such a study would be most interesting.

\begin{table*}
\caption{Merger rate of intermediate mass close binaries and breakdown into systems that merge during stable Roche lobe overflow, during common envelope evolution, and during spiral-in with a White Dwarf.}
\label{table2}
\centering
\begin{tabular}{c c c c c c c c}
\hline
$q$-distr. & $\beta$ & $\alpha\lambda$ & AM & Merger & during & during & during spiral-in with \\
 & & & loss & rate & Roche lobe overflow & common envelope & White Dwarf \\
\hline
flat & 1.0 & 1.0 & $L_2(\eta=2.3)$ & 76\% & 6\% & 30\% & 65\%\\
flat & 0.5 & 1.0 & $L_2(\eta=2.3)$ & 86\% & 23\% & 30\% & 48\%\\
flat & 0.0 & 1.0 & $L_2(\eta=2.3)$ & 88\% & 34\% & 30\% & 37\%\\
flat & 1.0 & 0.5 & $L_2(\eta=2.3)$ & 80\% & 6\% & 41\% & 54\%\\
flat & 0.0 & 0.5 & $L_2(\eta=2.3)$ & 90\% & 34\% & 41\% & 26\%\\
flat & 1.0 & 0.1 & $L_2(\eta=2.3)$ & 83\% & 6\% & 57\% & 37\%\\
flat & 0.5 & 0.1 & $L_2(\eta=2.3)$ & 91\% & 23\% & 57\% & 20\%\\
flat & 1.0 & "Dewi" & $L_2(\eta=2.3)$ & 82\% & 6\% & 46\% & 49\%\\
flat & 1.0 & 1.0 & O & 74\% & 0\% & 30\% & 70\%\\
flat & 0.5 & 1.0 & O & 70\% & 0\% & 30\% & 70\%\\
flat & 1.0 & 1.0 & $L_2(\eta=1)$ & 76\% & 3\% & 30\% & 67\%\\
flat & 0.5 & 1.0 & $L_2(\eta=1)$ & 83\% & 7\% & 30\% & 64\%\\
Garmany & 1.0 & 1.0 & $L_2(\eta=2.3)$ & 73\% & 4\% & 27\% & 69\%\\
Garmany & 0.0 & 1.0 & $L_2(\eta=2.3)$ & 89\% & 38\% & 27\% & 36\%\\
Hogeveen & 1.0 & 1.0 & $L_2(\eta=2.3)$ & 81\% & 11\% & 35\% & 54\%\\
Hogeveen & 0.0 & 1.0 & $L_2(\eta=2.3)$ & 85\% & 26\% & 35\% & 39\%\\
\hline
\end{tabular}
\end{table*}

\section{Application to Globular Clusters}

Massive AGBs of a first generation of stars in Globular Clusters are by now the only stars that successfully explain the observed anticorrelation of Mg-Al in second generation stars but to explain the observed anticorrelation of O-Na in these second generation stars, a dilution model is needed where AGB matter mixes with pristine gas or gas with a pristine chemical composition as far as Mg. Al, O and Na are concerned. How do intermediate mass close binaries influence this self-pollution scenario? The first effect is obvious: the higher the frequency of intermediate mass stars in intermediate mass close binaries the lower the frequency of single stars that will make it to the AGB. As an example, in a Globular Cluster with 100\% single stars and a Kroupa IMF, 9-10\% of the total mass of a population of first generation stars is contained in the envelopes of single AGB stars. However, in a Globular Cluster where 50\% of the total mass of the intermediate mass star range is contained in intermediate mass close binaries (for a flat mass ratio distribution this means an intermediate mass close binary frequency by number of 33\%), only 4-5\% of the total mass is contained in the envelopes of single AGB stars. The second effect: first generation intermediate mass close binaries may provide in a natural way the gas with a pristine chemical composition that is needed to explain the O-Na anti-correlation that is observed in Globular Clusters. This can be illustrated as follows: suppose that 50\% of the total mass of the intermediate mass star range are intermediate mass close binaries with a period less than 10 years. 4-5\% of the total mass of a first generation is contained in the envelopes of single AGB stars that is affected by hot bottom burning. Less than 1\% of the total mass is contained in the envelopes of binary components that is affected by hot bottom burning. 4-5\% of the total mass is lost by intermediate mass close binaries, it is He-enriched but as far as Na, O, Mg and Al is concerned it has a pristine chemical composition. Thus, the total amount of AGB mass that is affected by hot bottom burning is diluted with a similar amount of mass that is He-enriched but not affected by hot bottom burning, and when we compare this result with the Globular Cluster simulations of D'Ercole et al (2010) this resembles what is needed in order to explain the O-Na anti-correlation. We do not discuss the mass budget in more detail because as shown by D'Ercole et al. (2008) and Vesperini et al. (2010), this requires a full dynamical study of the dense stellar system capable to determine the fraction of first generation stars and binaries that escape from the cluster, and this is far beyond the scope of the present paper.

\section{Conclusions}

Even accounting for the uncertainties of the process of Hot Bottom Burning during the AGB phase of single stars, the AGB self-pollution scenario explains many observations of the younger generation of stars in Globular Clusters. Some important issues still need to be clarified, and one of them is the observed anticorrelations of Na-O and Al-Mg. The latter can be understood if matter that has been affected by the Hot Bottom Burning during the AGB phase in single stars is mixed with matter that has not been affected. Where this unaffected matter comes from is still a matter of debate but we have shown in the present paper that when a significant number of intermediate mass stars are born in close binaries, then the binary mass loss phases which are related to the Roche lobe overflow and/or common envelope  process can provide in a natural way the unaffected matter that is needed to explain the anticorrelations.  

\begin{acknowledgements}
We thank the anonymous referee for very helpful comments and suggestions.
\end{acknowledgements}

\end{document}